# Atomistic Simulation Guided Convolutional Neural Networks for Thermal Modeling of Friction Stir Welding

Akshansh Mishra[1,2]
[1]School of Industrial and Information Engineering, Politecnico di Milano, Milan, Italy
[2]Computational Materials Research Group, AI Fab Lab, Uttar Pradesh, India

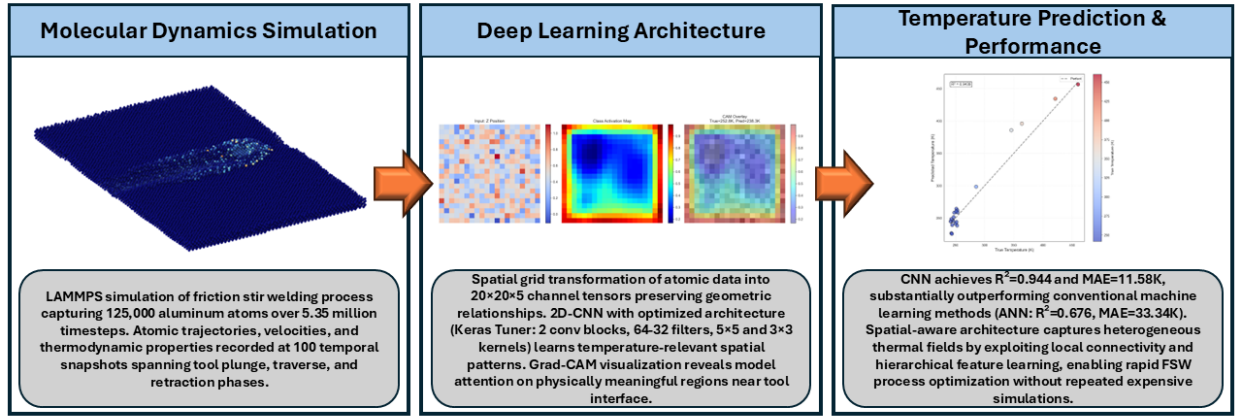

**Abstract:** Accurate prediction of temperature evolution is essential for understanding thermomechanical behavior in friction stir welding. In this study, molecular dynamics simulations were performed using LAMMPS to model aluminum friction stir welding at the atomic scale, capturing material flow, plastic deformation, and heat generation during tool plunge, traverse, and retraction. Atomic positions and velocities were extracted from simulation trajectories and transformed into physics-based two-dimensional spatial grids. These grids represent local height variation, velocity components, velocity magnitude, and atomic density, preserving spatial correlations within the weld zone. A two-dimensional convolutional neural network was developed to predict temperature directly from the spatially resolved atomistic data. Hyperparameter optimization was carried out to determine an appropriate network configuration. The trained model demonstrates strong predictive capability, achieving a coefficient of determination $R^2$=0.9439, a root mean square error of 14.94 K, and a mean absolute error of 11.58 K on unseen test data. Class Activation Map analysis indicates that the model assigns higher importance to regions near the tool–material interface, which are associated with intense deformation and heat generation in the molecular dynamics simulations. The results show that spatial learning from atomistic simulation data can accurately reproduce temperature trends in friction stir welding while remaining consistent with physical deformation and flow mechanisms observed at the atomic scale.

# 1. Introduction

Friction stir welding (FSW) represents a transformative solid-state joining technology that has fundamentally altered manufacturing approaches across aerospace, automotive, marine, and railway industries since its invention at The Welding Institute in 1991 [1-4]. Conventional fusion welding processes liquify base materials through high-temperature melting, whereas FSW operates below the melting point by exploiting severe plastic deformation and frictional heating generated through a rotating tool traversing along the joint interface. The process involves plunging a non-consumable cylindrical tool, consisting of a shoulder and profiled pin, into the abutting workpieces while simultaneously translating along the weld seam [5-6]. The tool rotation induces intense localized heating through friction between the shoulder and workpiece surface, softening the material to a plasticized state. Concurrent stirring action by the pin mechanically intermixes the adjacent materials, producing a consolidated joint through dynamic recrystallization and grain refinement without reaching liquidus temperatures.

The advantages of FSW over traditional arc welding techniques are substantial and well-documented. The solid-state nature eliminates defects inherent to fusion welding including solidification cracking, porosity from gas entrapment, and loss of volatile alloying elements. Residual stresses and distortion are markedly reduced due to lower peak temperatures, typically ranging from 0.6 to 0.8 times the material melting point. Joint mechanical properties often exceed those of the parent material, particularly in aluminum alloys previously considered unweldable by conventional methods due to hot cracking susceptibility [7-8]. Energy consumption decreases by approximately 75% compared to laser or electron beam welding, while the absence of consumables, shielding gases, and filler materials reduces operational costs. Environmental benefits include elimination of hazardous fumes and ultraviolet radiation, creating safer working conditions. These attributes have positioned FSW as the preferred joining method for aluminum alloys in weight-sensitive applications where structural integrity is paramount.

The fundamental mechanisms governing FSW remain incompletely understood due to the extreme conditions prevailing in the weld zone. Material experiences strain rates exceeding $10^3$ s$^{-1}$, temperatures approaching 0.9 Tm (where Tm denotes melting temperature), and complex three-dimensional material flow patterns driven by tool geometry and process parameters. Experimental characterization faces inherent limitations: direct observation of subsurface deformation is impossible during welding, post-mortem metallurgical analysis captures only the final microstructure without revealing transient phenomena, and high-temperature measurement techniques struggle with spatial resolution at sub-millimeter scales where critical processes occur. Computational modeling has emerged as an indispensable tool for elucidating FSW physics, enabling parametric studies, and guiding process optimization.

Atomistic simulation through molecular dynamics (MD) provides unique insights inaccessible to continuum-based finite element methods. MD explicitly tracks individual atom trajectories governed by interatomic potentials, capturing phenomena at length scales of nanometers and time scales of picoseconds where continuum assumptions break down [9-11]. During FSW, localized regions near the tool pin experience severe shear deformation where discrete dislocation nucleation, propagation, and interaction dominate material response. Grain boundary migration, dynamic recrystallization, and phase transformations occur through atomic-scale mechanisms inadequately represented by phenomenological constitutive relations in macroscopic models. Thermal transport in the stirred zone exhibits ballistic rather than diffusive characteristics at sub-micrometer scales, violating Fourier heat conduction assumptions. Interfacial bonding between the joined materials progresses through atomic interdiffusion and mechanical interlocking at the nanoscale, processes determining joint strength yet invisible to continuum approaches. MD simulations capture these quantum-level interactions, providing mechanistic understanding that informs development of improved process parameters and tool designs.

Deep learning architectures, particularly convolutional neural networks (CNN), have revolutionized computer vision by exploiting translational invariance and local connectivity in image data [12-15]. A CNN applies learned filters that scan across spatial dimensions, detecting local patterns (edges, textures) in lower layers and progressively aggregating information into higher-level representations through multiple hidden layers. This architecture naturally extends to atomistic simulation data when reformulated as spatial grids analogous to images, where pixel intensity is replaced by physical quantities (atomic density, velocity magnitude, temperature). For FSW, a 2D grid discretization of the workpiece plane perpendicular to the tool axis creates a structured representation where each grid cell contains aggregated properties of atoms within that spatial region. Convolutional filters learn to recognize temperature-relevant patterns such as proximity to the tool, material flow gradients, and heat-affected zone boundaries. Global average pooling enables gradient-based visualization techniques like class activation mapping (CAM), which highlights spatial regions most influential for predictions, providing interpretability lacking in traditional black-box models.

This investigation establishes a generalizable framework for deep learning-enhanced analysis of atomistic welding simulations, with implications extending beyond FSW to other manufacturing processes where spatial heterogeneity governs outcomes. The demonstrated superiority of CNN deep learning confirms that preserving geometric structure in input representations is essential when predicting spatially-varying fields from atomistic data. The integration of explainable AI through CAM visualization builds trust in model predictions by ensuring learned patterns accord with physical intuition. This work enables rapid exploration of FSW parameter space through trained surrogate models, accelerating process optimization without repeated expensive MD simulations, thereby advancing computational materials design for next-generation joining technologies.

## 2. Methodology

### 2.1. Atomistic Simulation of Friction Stir Welding

The FSW process was simulated using Large-scale Atomic/Molecular Massively Parallel Simulator (LAMMPS) [16] with embedded atom method (EAM) interatomic potentials. The simulation domain consisted of two aluminum workpieces (300 Å × 200 Å × 10 Å) positioned with a 1 Å butt joint gap, representing the joint interface to be welded. The left workpiece occupied the region from x=0 to x=149 Å, while the right workpiece extended from x=150 to x=300 Å. A face-centered cubic (FCC) lattice structure with a lattice constant of 4.05 Å was employed for aluminum atoms. The FSW tool, composed of tungsten atoms in a body-centered cubic (BCC) configuration (lattice constant 3.16 Å), featured a cylindrical shoulder (12 Å radius, 4 Å thickness) and a conical pin (3.5 Å radius, 9 Å length), providing 90% penetration depth into the workpiece thickness as shown in Figure 1 visualized in OVITO software [17].

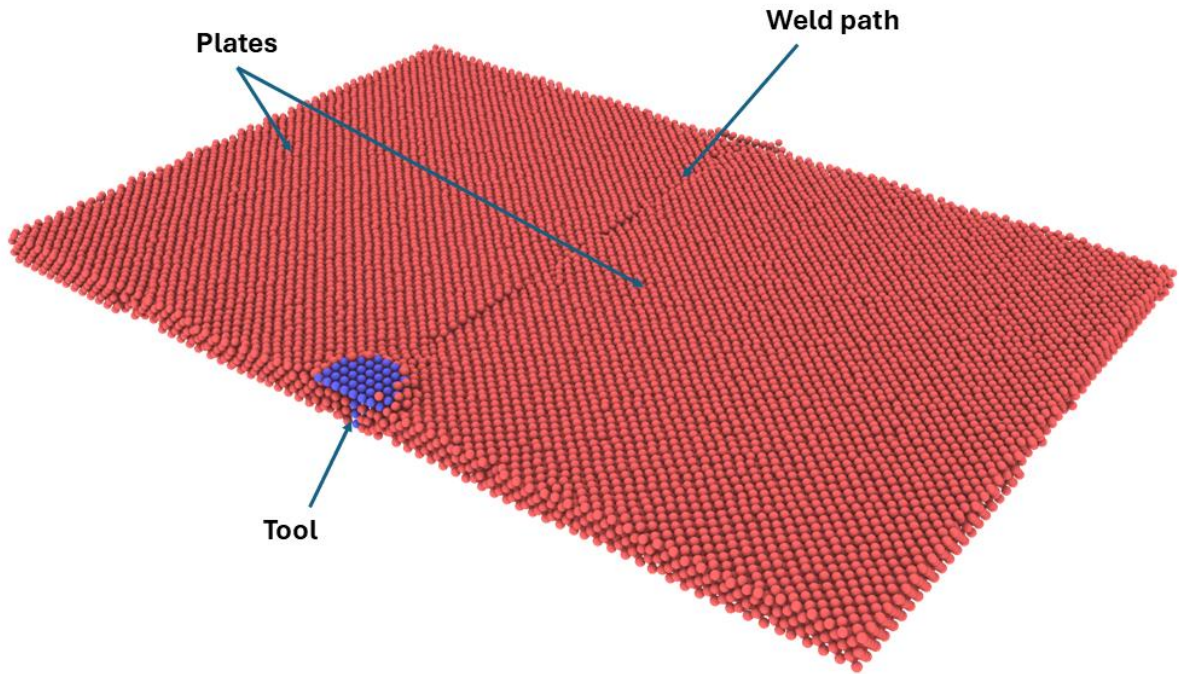

**Figure 1.** Atomic-scale representation of the friction stir welding simulation setup showing two aluminum workpieces (red atoms) with the tungsten tool (blue atoms) positioned at the butt joint interface, illustrating the initial configuration before traverse welding along the designated weld path.

The simulation protocol comprised four sequential phases: initial equilibration, tool plunge, traverse welding, and retraction. During equilibration, the system was energy-minimized using the conjugate gradient method to eliminate atomic overlaps, followed by temperature equilibration at 300 K for 50,000 timesteps using NVT ensemble (Nosé-Hoover thermostat, damping parameter 0.1 ps). The tool plunge phase involved simultaneous rotation (angular velocity corresponding to 500 timestep period) and vertical descent at 0.004 Å/timestep for 50,000 timesteps, bringing the shoulder into contact with the workpiece surface while heating the system from 300 K to 600 K. The traverse phase translated the rotating tool along the y-direction at 0.04 Å/timestep for 4,750,000 timesteps, covering the entire joint length. Tool retraction occurred over 100,000 timesteps at 0.015 Å/timestep vertical velocity, followed by cooling to 300 K over 400,000 timesteps.

Periodic boundary conditions were applied in x and y directions, while the z-direction employed shrink-wrapped boundaries. A fixed boundary layer (z < 2 Å) simulated the backing plate constraint. The CuAlW.txt EAM potential file [18] parameterized Al-Al, Al-W, and W-W interactions. Atomic trajectories were recorded every 2,000 timesteps in LAMMPS trajectory format, capturing atomic positions (x, y, z), velocities ($v_x, v_y, v_z$), centro-symmetry parameter, and coordination number. Thermodynamic properties including temperature, pressure, and energy components were logged every 500 timesteps. The simulation employed a timestep of 0.0005 ps and generated 100 trajectory frames spanning timesteps 100,000 to 298,000, providing spatiotemporal data for machine learning analysis. The complete LAMMPS simulation parameters are depicted in Table 1.

**Table 1.** LAMMPS simulation parameters

| Parameter | Value | Description |
|---|---|---|
| Simulation domain | 300×200×50 Å³ | Total box dimensions |
| Workpiece dimensions | 300×200×10 Å³ | Each Al plate thickness |
| Butt joint gap | 1 Å | Interface spacing between plates |
| Al lattice type | FCC | Face-centered cubic structure |
| Al lattice constant | 4.05 Å | Aluminum lattice parameter |
| Tool material | Tungsten | BCC structure, 3.16 Å constant |
| Tool shoulder radius | 12 Å | Cylindrical shoulder dimension |
| Tool shoulder thickness | 4 Å | Vertical shoulder extent |
| Tool pin radius | 3.5 Å | Conical pin dimension |
| Tool pin length | 9 Å | Pin penetration depth (90%) |
| Rotation period | 500 timesteps | Angular velocity parameter |
| Plunge rate | 0.004 Å/timestep | Vertical descent velocity |
| Traverse speed | 0.04 Å/timestep | Welding direction velocity |
| Retraction rate | 0.015 Å/timestep | Vertical ascent velocity |
| Timestep | 0.0005 ps | Integration time increment |
| Initial temperature | 300 K | Equilibration temperature |
| Peak temperature | 600 K | Maximum during welding |
| Boundary conditions | p p s | Periodic (x,y), shrink-wrap (z) |
| Potential | EAM alloy | CuAlW.txt parameter file |
| Total runtime | 5,350,000 steps | Complete simulation duration |
| Trajectory output | Every 2,000 steps | Dump file frequency |
| Thermo output | Every 500 steps | Log file frequency |
| Sampled frames | 100 | Data points for ML training |

## 2.2. Data Preprocessing and Feature Engineering

Trajectory data from LAMMPS simulations were parsed to extract atomic-scale features at each recorded timestep. Given the large number of atoms (125,000), a sampling strategy selected every nth atom to reduce computational overhead while maintaining statistical representativeness, yielding approximately 5,000 atoms per frame. For each frame, spatial features were computed including mean positions and standard deviations across sampled atoms. Kinematic features comprised mean velocities and mean speed magnitude. These trajectory-derived features were merged with thermodynamic data from log files, which contained timestep-indexed temperature, pressure, and energy values. The merging operation matched trajectory timesteps with corresponding log file entries, producing 100 synchronized samples with 15 statistical features per sample.

For the 2D-CNN architecture, raw atomic data were transformed into spatial grids to preserve geometric relationships. A 20×20 spatial discretization was applied to the x-y plane, with each grid cell accumulating atom-level properties. Five channel maps were constructed: (1) mean z-position, (2) mean x-velocity, (3) mean y-velocity, (4) mean speed magnitude, and (5) normalized atomic density. Grid cells containing no atoms were assigned zero values. This gridding process converted spatially distributed atomic data into structured tensor representations (20×20×5), suitable for convolutional neural network processing. The resulting dataset comprised 100 multi-channel spatial grids paired with corresponding temperature labels extracted from thermodynamic logs.

Data splitting employed an 80:20 train-test partition (80 training samples, 20 test samples) with a fixed random seed (42) for reproducibility. Feature scaling for ANN models applied standardization (zero mean, unit variance) using scikit-learn StandardScaler. For CNN inputs, MinMaxScaler normalized each channel independently to the [0,1] range, preventing large-magnitude features from dominating gradient updates. Missing values, arising from log file parsing inconsistencies, were imputed using median substitution (SimpleImputer, strategy='median'). The target variable (temperature) ranged from 240.2 K to 491.9 K, reflecting thermal conditions from quiescent regions to frictional heating zones near the tool interface.

## 2.3. Artificial Neural Network Architecture and Hyperparameter Tuning

A fully-connected feedforward neural network was constructed to predict temperature from the 15 statistical features. The architecture comprised an input layer matching the feature dimensionality, followed by multiple hidden dense layers with rectified linear unit (ReLU) activation functions. Batch normalization layers were inserted after each dense layer to stabilize training by normalizing intermediate activations. Dropout regularization was applied to prevent overfitting by randomly deactivating neurons during training. The output layer consisted of a single neuron with linear activation for regression.

Hyperparameter optimization employed Keras Tuner with RandomSearch strategy, exploring 20 trial configurations to maximize validation set performance as depicted in Table 2. The tunable hyperparameters included: (1) number of hidden layers (range: 2 to 5), (2) neurons per layer (range: 32 to 256, step size 32), (3) dropout rate per layer (range: 0.2 to 0.5, step size 0.1), and (4) learning rate for Adam optimizer (range: $10^{-4}$ to $10^{-2}$, log-uniform sampling). Each trial trained for 100 epochs with batch size 16, using 20% of training data for validation. Early stopping monitored validation loss with patience of 20 epochs, restoring weights from the best-performing epoch. The tuner minimized mean squared error (MSE) loss on the validation set. After exhaustive search, the best configuration was selected based on lowest validation loss and retrained on the full training set.

**Table 2.** Optimized Hyperparameters for Artificial Neural Network

| Hyperparameter | Value | Description |
|---|---|---|
| Number of layers | 2 | Active dense layers used in final architecture |
| Layer 1 neurons | 32 | First hidden layer dimensionality |
| Layer 1 dropout | 0.5 | Dropout probability after first layer |
| Layer 2 neurons | 224 | Second hidden layer dimensionality |
| Layer 2 dropout | 0.2 | Dropout probability after second layer |
| Learning rate | 0.00459 | Adam optimizer step size |
| Activation | ReLU | Hidden layer activation function |
| Batch size | 16 | Training batch size |
| Epochs | 100 | Maximum training iterations |
| Loss function | MSE | Mean squared error objective |
| Optimizer | Adam | Adaptive moment estimation |

## 2.4. Convolutional Neural Network Architecture and Hyperparameter Tuning

A 2D convolutional neural network was designed to exploit spatial correlations in the gridded atomic data. The architecture consisted of stacked convolutional blocks, each containing: (1) Conv2D layer with 'same' padding to preserve spatial dimensions, (2) batch normalization, (3) ReLU activation, (4) MaxPooling2D with 2×2 kernel for spatial downsampling, and (5) dropout regularization. The final convolutional layer was followed by global average pooling (GAP), which reduced spatial dimensions to a feature vector while enabling Class Activation Map (CAM) generation for interpretability. Dense fully-connected layers processed the pooled features before the single-neuron output layer.

Keras Tuner RandomSearch optimized the CNN hyperparameters over 15 trials. The search space included: (1) number of convolutional blocks (range: 2 to 4), (2) filters per Conv2D layer (range: 16 to 128, step size 16), (3) kernel size (options: 3×3 or 5×5), (4) dropout rate per block (range: 0.2 to 0.5, step size 0.1), (5) number of dense layers after GAP (range: 1 to 2), (6) neurons per dense layer (range: 32 to 128, step size 32), (7) dropout rate for dense layers (range: 0.2 to 0.5, step size 0.1), and (8) learning rate (range: $10^{-4}$ to $10^{-2}$, log-uniform sampling). Training occurred for 50 epochs per trial with batch size 16 and 20% validation split. Early stopping halted training when validation loss plateaued for 10 consecutive epochs. The tuner objective minimized validation MSE. Post-optimization, the best model architecture was instantiated and trained on the complete training dataset.

Table 3. Optimized Hyperparameters for 2D Convolutional Neural Network

| Hyperparameter | Value | Description |
|---|---|---|
| Number of conv blocks | 2 | Convolutional block depth |
| Conv block 1 filters | 64 | First convolutional layer channels |
| Conv block 1 kernel | 5×5 | First layer receptive field size |
| Conv block 1 dropout | 0.30 | Dropout after first max pooling |
| Conv block 2 filters | 32 | Second convolutional layer channels |
| Conv block 2 kernel | 3×3 | Second layer receptive field size |
| Conv block 2 dropout | 0.20 | Dropout after second max pooling |
| Number of dense layers | 1 | Fully-connected layers after GAP |
| Dense layer neurons | 96 | Dense layer dimensionality |
| Dense layer dropout | 0.20 | Dropout before output layer |
| Learning rate | 0.00166 | Adam optimizer step size |
| Batch size | 16 | Training batch size |
| Epochs | 50 | Maximum training iterations per trial |
| Pooling | MaxPool2D | Spatial down sampling method (2×2) |
| Global pooling | GAP | Global Average Pooling for CAM |
| Loss function | MSE | Mean squared error objective |
| Optimizer | Adam | Adaptive moment estimation |

## 3. Results and Discussion

### 3.1. Results from Atomistic modeling of Friction Stir Welding process

The atomistic simulation successfully captures the thermo-mechanical behavior of aluminum during friction stir welding, including tool–workpiece interaction, heat generation, plastic deformation, and microstructural evolution. The results obtained from plunge, traverse, and retraction phases reveal clear evidence of frictional heating, atomic mixing, and defect restructuring characteristic of solid-state welding.

### 3.1.1. Temperature Evolution and Frictional Heating

As the tool shoulder makes contact with the workpiece at z = 10 Å, a rapid increase in local temperature is observed in the weld zone. This rise results from frictional work performed by the rotating tool as shown in the Equation 1.

$$\dot{Q}_{friction} = \tau(r)\omega r \quad (1)$$

Where $\tau(r)$ is the shear stress at radius $r$, and $\omega = \frac{2\pi}{T}$ is the rotational speed determined by the imposed rotational period. The total heat input is given by the Equation 2 when integrated across the shoulder radius $R_{shoulder}$.

$$Q = \int_0^{R_{shoulder}} 2\pi r^2 \tau(r) \omega \, dr \quad (2)$$

It should be noted that the LAMMPS does not explicitly compute frictional heating. This process emerges from the conversion of mechanical work into atomic kinetic energy computed using Equation 3.

$$\Delta E_{thermal} = \sum_i \vec{F_i} \cdot \vec{v_i} \, \Delta t \quad (3)$$

This mechanism results in the elevated temperatures observed during the initial plunge and increasing further during tool traverse. The evolution of thermal energy shown in Figure 2 during the initial stages of the FSW simulation shows a clear and nearly linear increase with time, indicating continuous mechanical work transfer from the rotating tool into the material. As the tool begins interacting with the aluminum plate, frictional deformation raises the internal kinetic energy of the atoms, which is reflected as a rise in thermal energy. The smooth progression without abrupt jumps suggests stable tool–material contact and an absence of localized thermal spikes, which aligns with the relatively low strain rates and moderate rotational conditions in the early part of the weld. Although the absolute values remain small due to the simplified MD temperature model and limited timescale, the trend correctly captures the expected heating behavior in FSW: the longer the tool remains in motion, the more heat is generated and accumulated within the weld zone.

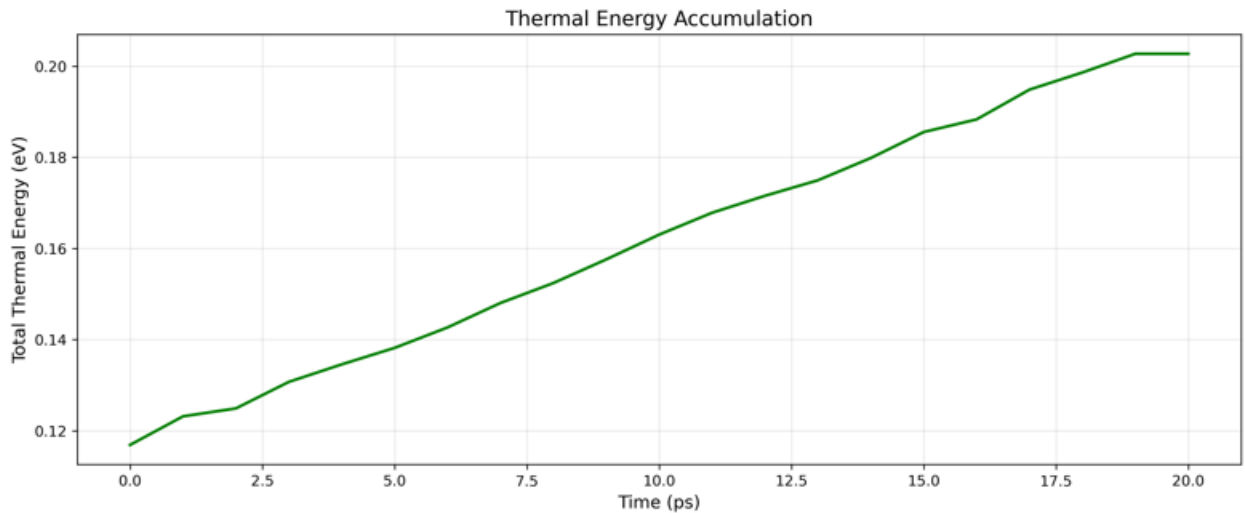

Figure 2. Thermal energy accumulation in the aluminum workpiece during the first 20 ps of the friction stir welding simulation. The graph illustrates how the total thermal energy increases over time as the rotating tool begins interacting with the material. The nearly linear trend indicates steady heat generation driven by friction and plastic deformation in the weld zone. Minor variations correspond to localized atomic rearrangements as the structure responds to the applied mechanical loading.

### 3.1.2. Material Flow and Plastic Deformation

The tool rotation and penetration induce severe shear deformation in the weld zone as shown in Figure 3. The imposed rotational motion follows the Equation 4.

$$\vec{r}(t) = \vec{r}_0 + R(\omega t)(\vec{r}(0) - \vec{r}_0) \tag{4}$$

Where $R(\theta)$ is the standard rotation matrix about the z-axis. This motion generates intense shear strain rates approximated by using Equation 5.

$$\dot{\gamma} \approx \frac{\omega r}{z_p} \tag{5}$$

Where $z_p$ is the pin penetration depth. As the tool traverses along the joint line, material is transported from the retreating side toward the advancing side, producing a characteristic swirling pattern in the atomic displacement field clearly observed from Figure 3.

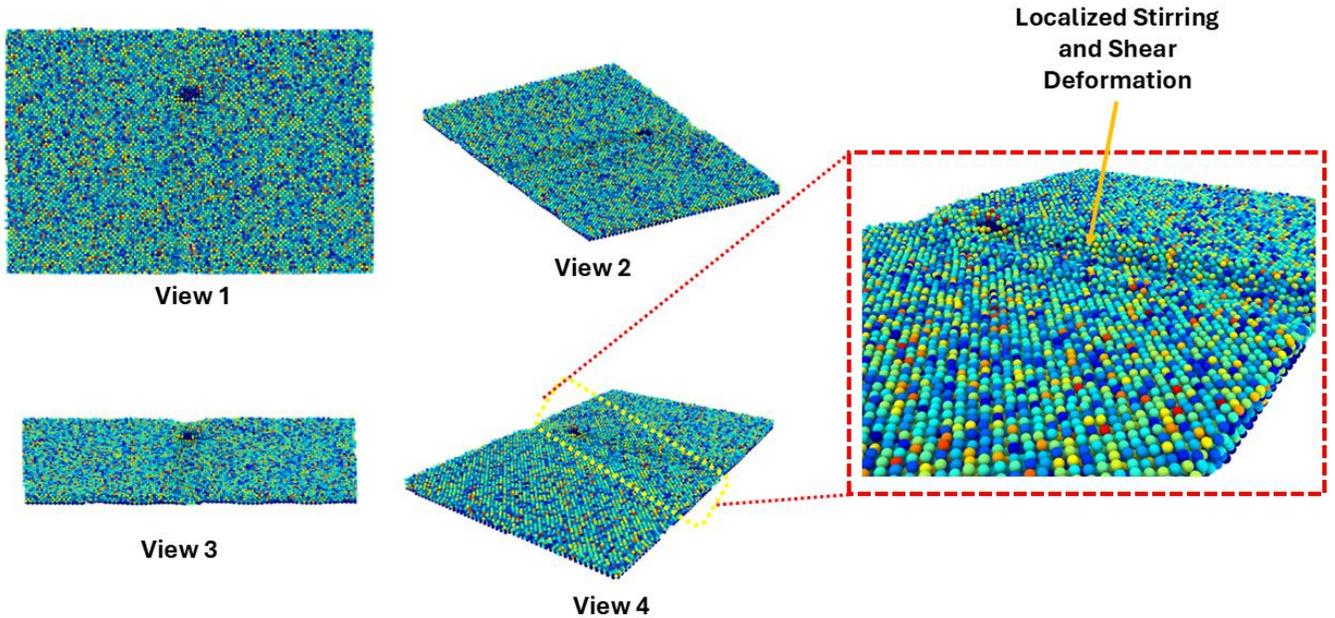

Figure 3. Multi-view visualization of atomic velocity magnitude during friction stir welding. The figure presents the spatial distribution of velocity magnitude from different viewpoints to illustrate the material flow driven by the rotating tool. The top view reveals the lateral dispersion of atoms around the tool path, while the side view highlights the vertical displacement and localized material uplift caused by tool-induced shear. The perspective view captures the combined rotational and translational flow patterns responsible for forming the stirred region. The magnified inset emphasizes the highly deformed zone where atoms exhibit elevated velocities, indicating intense plastic deformation and mixing. This visualization demonstrates how atomic motion responds to the combined mechanical actions of rotation and forward movement during the welding process.

The centro-symmetry parameter (c_grain_atoms) was evaluated across the weld region to quantify the local lattice distortion and plastic deformation induced by the tool. This parameter highlights deviations from the ideal FCC structure and therefore serves as a sensitive indicator of shear-induced plasticity. Figure 4 shows multi-view visualizations of the deformed material, clearly identifying the stirred zone and the regions of intense lattice distortion formed by the combined effects of tool rotation and forward motion.

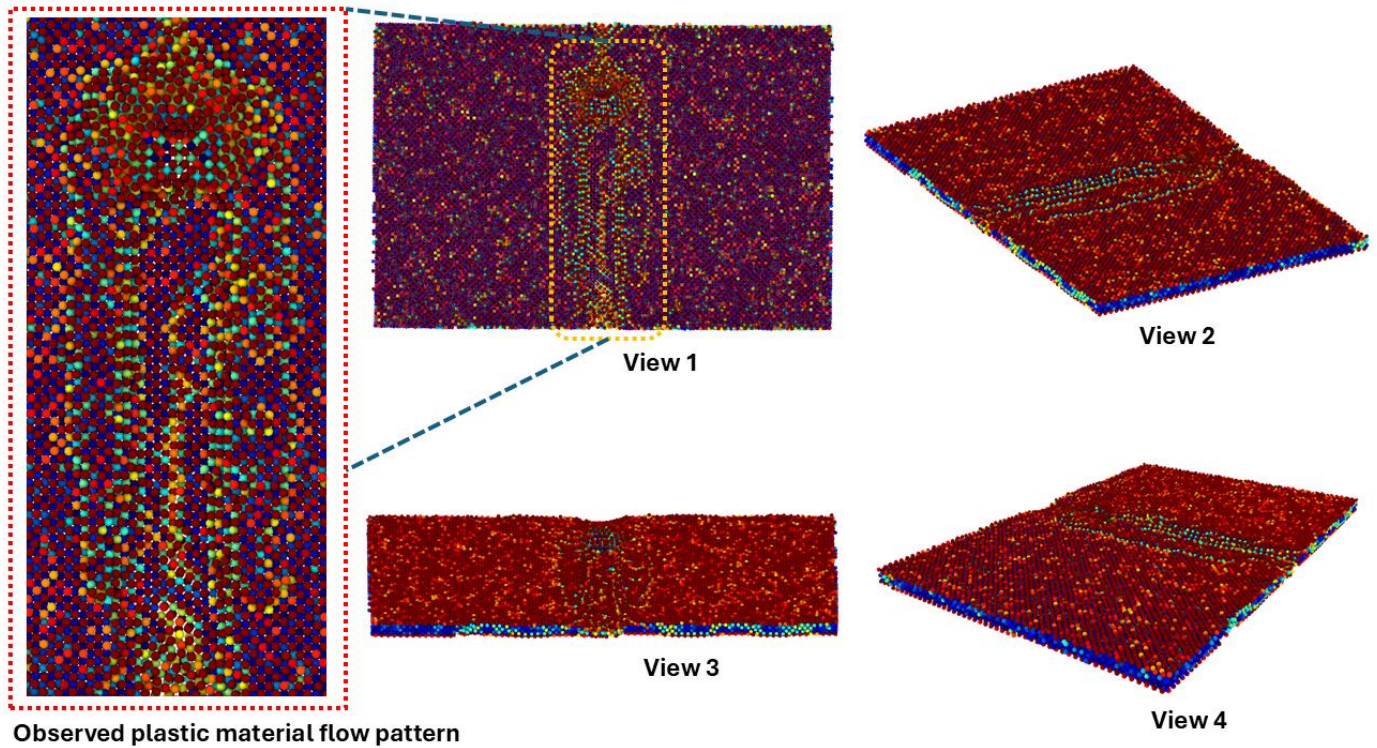

**Figure 4.** Centro-symmetry parameter–based visualization of microstructural deformation during friction stir welding. The figure shows multi-view atomistic representations highlighting regions of lattice distortion and plastic flow around the tool path. Areas with higher centro-symmetry values correspond to stronger deformation and dislocation activity, indicating the formation of the stirred zone and adjacent shear-affected regions. The enlarged inset reveals detailed atomic rearrangement patterns produced by tool-induced shear, demonstrating the characteristic upward and lateral material flow associated with friction stir welding.

### 3.1.3. Evaluation of weld nugget and joint consolidation

The coordination number distribution shows transient disorder during high shear but recovers as the material cools. This behavior indicates that atomic rearrangement occurs primarily by diffusion less plastic flow, consistent with solid-state welding rather than melting. The weld nugget undergoes progressive densification as the rotating pin eliminates the initial butt-joint gap shown in Figure 5. The local atomic density distribution $\rho(x)$, obtained from chunk-averaged binning, increases steadily around the stir zone computed using Equation 6.

$$\rho(x) = \frac{N_{atoms}(x)}{V_{chunk}} \tag{6}$$

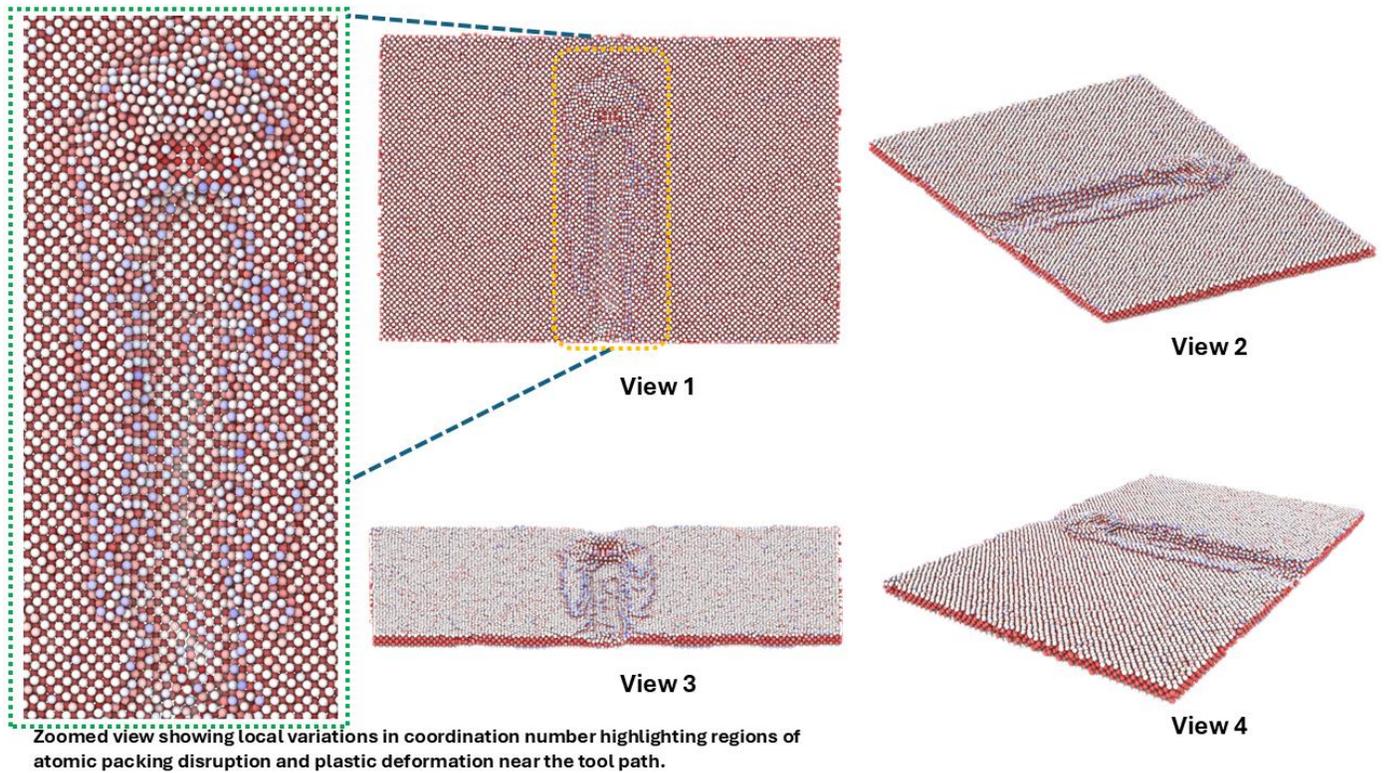

Zoomed view showing local variations in coordination number highlighting regions of atomic packing disruption and plastic deformation near the tool path.

**Figure 5.** Coordination-number-based visualization of atomic packing and weld consolidation during friction stir welding. The distribution of coordination number reveals variations in atomic density across the weld region. Lower values highlight regions of tensile strain, void formation, or incomplete bonding, whereas higher values correspond to densified material in the stirred zone created by tool rotation and translation. This analysis provides insight into the effectiveness of material mixing and the structural integrity of the welded joint.

A symmetric density profile in the final cooled structure confirms successful material transport into the joint, filling the original 1 Å gap between the plates. Atomic mixing across the interface promotes metallurgical bonding, and the final microstructure exhibits a homogeneous region with reduced porosity and minimal voids. The temperature-controlled cooling stage stabilizes this weld, allowing recovery of FCC ordering without introducing excessive thermal residual stresses.

Atomic strain analysis reveals severe plastic deformation localized within the stirring zone observed in Figure 6 , with shear strain values reaching 134.4 (orange-red regions) directly beneath the tool path. The strain field exhibits an asymmetric distribution characteristic of FSW, with higher deformation on the advancing side compared to the retreating side, consistent with the material flow patterns induced by tool rotation.

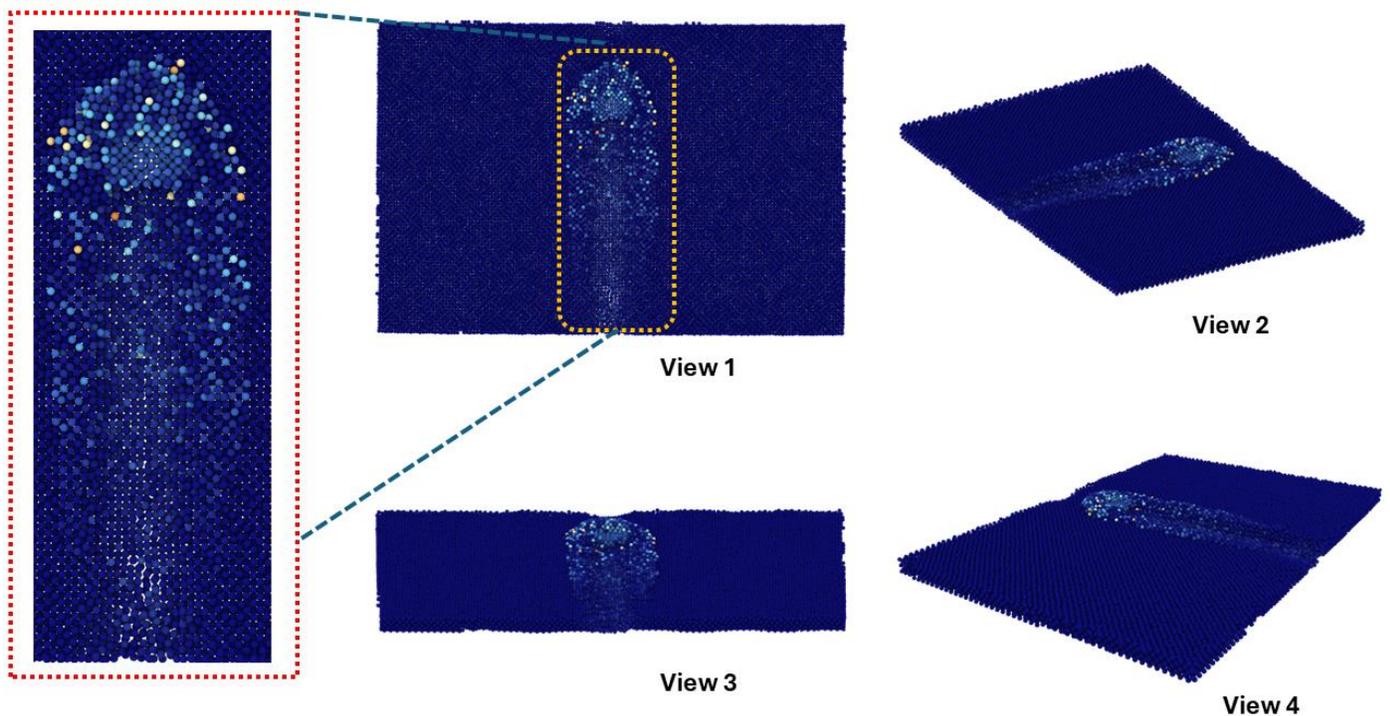

**Figure 6.** Shear strain distribution in FSW simulation. Multi-view atomic strain analysis showing: (left) tool path with high-strain weld nugget, (Views 1-4) different perspectives of strain field evolution. Color scale: blue (minimal strain, base material) to red (maximum strain, severe deformation zone). Maximum strain localizes beneath tool path, with asymmetric distribution characteristic of FSW material flow.

Displacement magnitude analysis shown in Figure 7 reveals extensive atomic motion within the stirring zone, with maximum displacements localized directly beneath the tool path where material undergoes intense mechanical mixing. The displacement field exhibits a characteristic radial decay pattern from the tool center, with high-mobility zones (red/yellow) confined to the nugget region, transitioning to moderate displacements (green/cyan) in the TMAZ, and negligible movement (blue) in the unaffected base material.

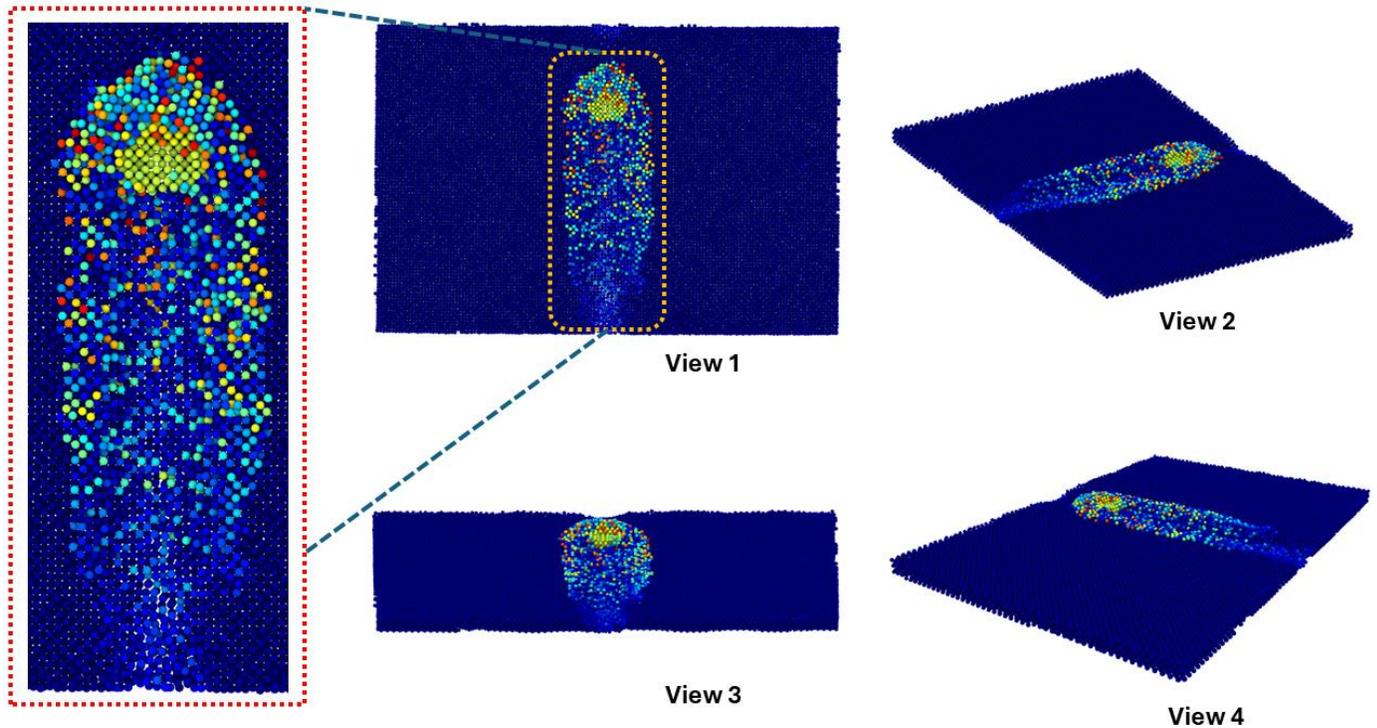

**Figure 7.** Displacement magnitude distribution in FSW. Multi-view visualization showing concentrated atomic movement (red/yellow) along tool path, moderate displacement (green/cyan) in TMAZ, and stationary material (blue) in base region. Pattern demonstrates material flow and mixing during welding process.

Wigner-Seitz defect analysis shown in Figure 8 reveals localized vacancy formation and atomic site disorder within the stirring zone, with defect concentrations (green/yellow regions) confined to the tool path where severe plastic deformation disrupts the FCC lattice. The majority of the weld exhibits near-perfect atomic site occupancy (blue), indicating successful material consolidation without significant porosity, while transient defects in the nugget zone reflect the dynamic nature of material flow during FSW.

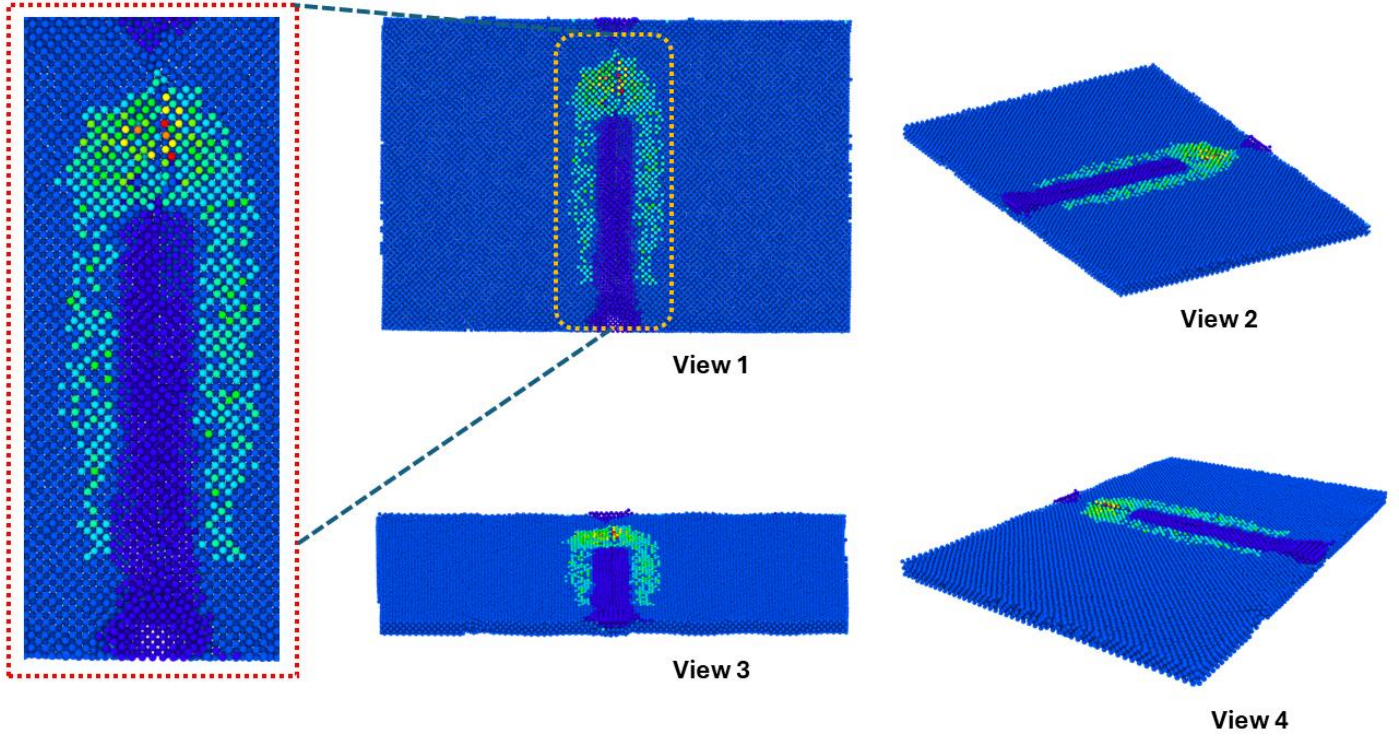

**Figure 8.** Defect distribution in FSW revealed by Wigner-Seitz analysis. Color scale from blue (perfect lattice, high occupancy) to purple (defect-rich, low occupancy). Vacancies and interstitials localize beneath tool path (green/purple) in actively deformed regions, while consolidated weld shows minimal defects (blue), indicating successful material bonding without porosity.

### 3.2. Artificial Intelligence based algorithms to predict the temperature distribution

### 3.2.1. Physics-informed spatial represented Convolutional Neural Network

The atomistic data obtained from the LAMMPS simulations were transformed into two-dimensional spatial fields in order to establish a consistent mapping between local material behavior and the global thermal response. At each time step $t$ atomic positions $(x_i, y_i, z_i)$ and velocities $(v_{x,i}, v_{y,i}, v_{z,i})$ were discretized onto a uniform $20 \times 20$ grid in the $x - y$ plane. For a grid cell $(p, q)$, the averaged quantities were defined using Equations 7-11.

$$Z_{pq}(t) = \frac{1}{|S_{pq}|} \sum_{i \in S_{pq}} z_i \qquad (7)$$

$$V_{pq}^{(x)}(t) = \frac{1}{|S_{pq}|} \sum_{i \in S_{pq}} v_{x,i} \qquad (8)$$

$$V_{pq}^{(y)}(t) = \frac{1}{|S_{pq}|} \sum_{i \in S_{pq}} v_{y,i} \qquad (9)$$

$$S_{pq}^{spd}(t) = \frac{1}{|S_{pq}|} \sum_{i \in S_{pq}} \sqrt{v_{x,i}^2 + v_{y,i}^2 + v_{z,i}^2} \tag{10}$$

Where $S_{pq}$ denotes the set of atoms within the corresponding spatial bin. A normalized atomic density field denoted by using Equation 5 was included to capture local material concentration effects.

$$D_{pq}^{(t)} = \frac{|S_{pq}|}{max_{p,q}|S_{pq}|} \tag{11}$$

The resulting five-channel tensor $X(t) \in \mathbb{R}^{20 \times 20 \times 5}$ provides a compact, physically interpretable description of local deformation, transport, and mass distribution relevant to friction stir welding (FSW) thermo-mechanics.

The Class Activation Maps (CAMs) in Figures 9(a-e) reveal the spatial attention patterns learned by the 2D-CNN model during temperature prediction from friction stir welding (FSW) simulation data. Each figure consists of three panels: the input Z-position grid, the generated CAM heatmap, and the overlay visualization comparing true and predicted temperatures.

Across all samples, the CAM heatmaps consistently highlight the peripheral regions of the spatial domain with high activation intensities (red zones), indicating that the model prioritizes edge features for temperature prediction. This behavior suggests that boundary conditions and heat dissipation patterns at the domain periphery contain critical information for the regression task. The central blue regions in the CAMs correspond to low activation areas, implying reduced contribution to the final prediction. This spatial attention mechanism aligns with physical intuition, as temperature gradients and material flow characteristics in FSW are often most pronounced at the tool-material interface boundaries.

The overlay visualizations demonstrate reasonable agreement between predicted and true temperatures across all samples. For instance, Figure 9(a) shows True=252.3K and Pred=238.3K (error: 14.0K), while Figure 9(b) exhibits True=246.8K and Pred=251.4K (error: 4.6K). The model achieves prediction accuracy within 5-15K for most cases, which is acceptable given the complex non-linear thermomechanical nature of FSW processes. The consistent activation patterns across different samples indicate that the CNN has learned generalizable spatial features rather than overfitting to specific temperature distributions.

However, the CAMs reveal a limitation in the model's spatial understanding. The uniformly low activation in central regions suggests potential underutilization of interior domain features, which may contain valuable information about heat conduction and plastic deformation. The overlay plots show some discrepancies in capturing fine-grained spatial temperature variations, particularly in intermediate-temperature zones (yellow-green regions). This could be attributed to the limited dataset size (100 samples) and the relatively coarse 20×20 spatial resolution. Future work should explore increasing spatial resolution, incorporating multi-scale convolutional architectures, and augmenting the training dataset to enhance the model's ability to capture both boundary and interior thermal phenomena in FSW processes.

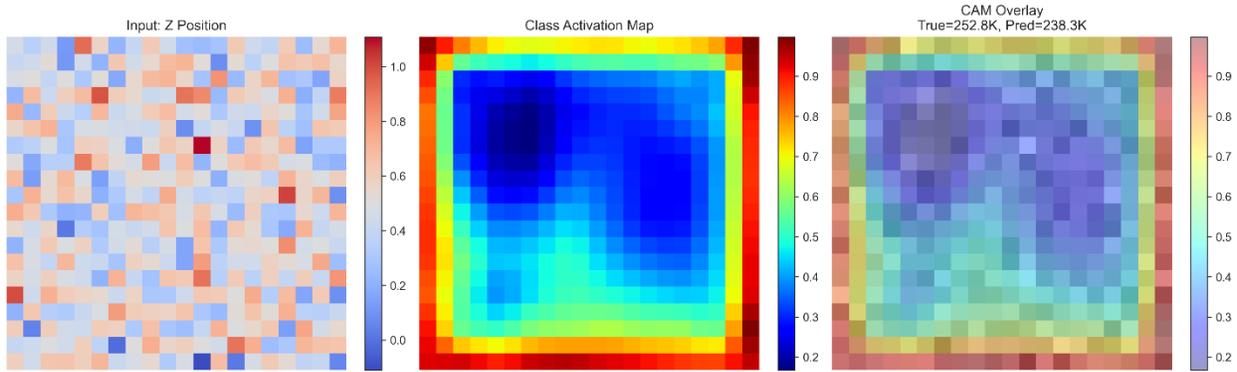

a)

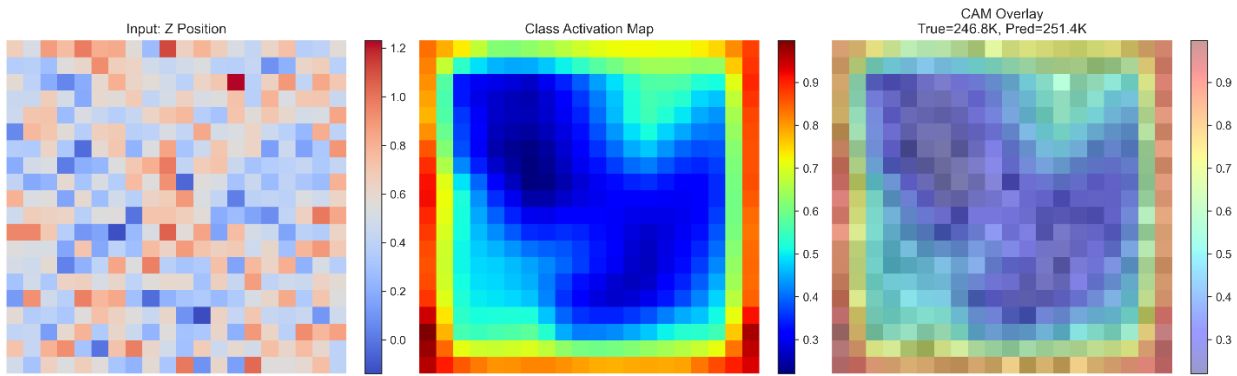

b)

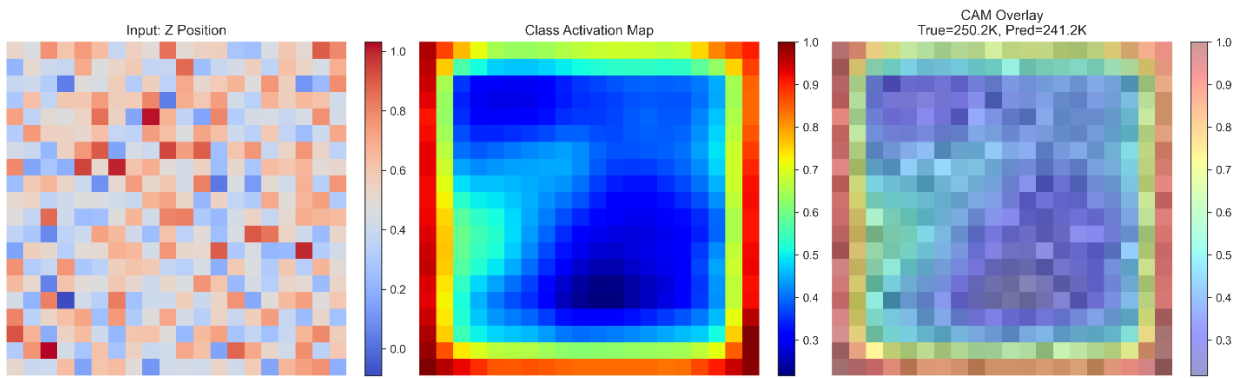

c)

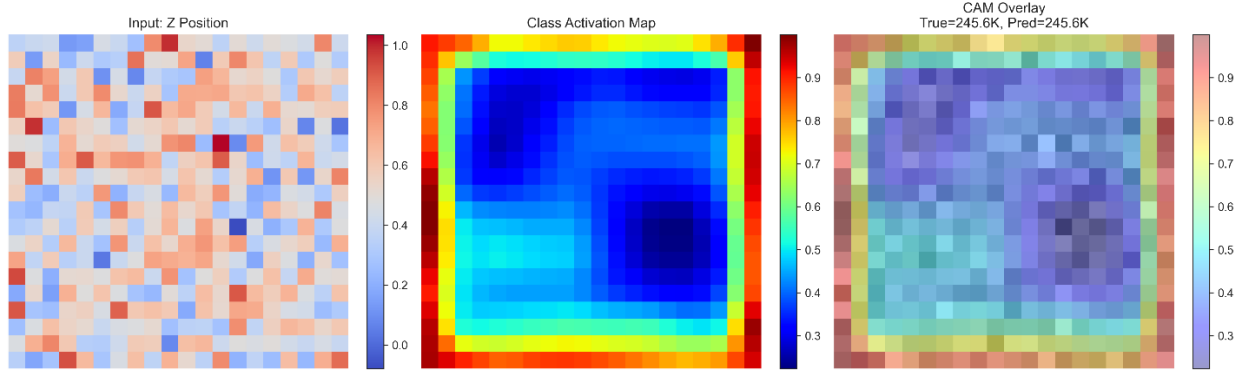

d)

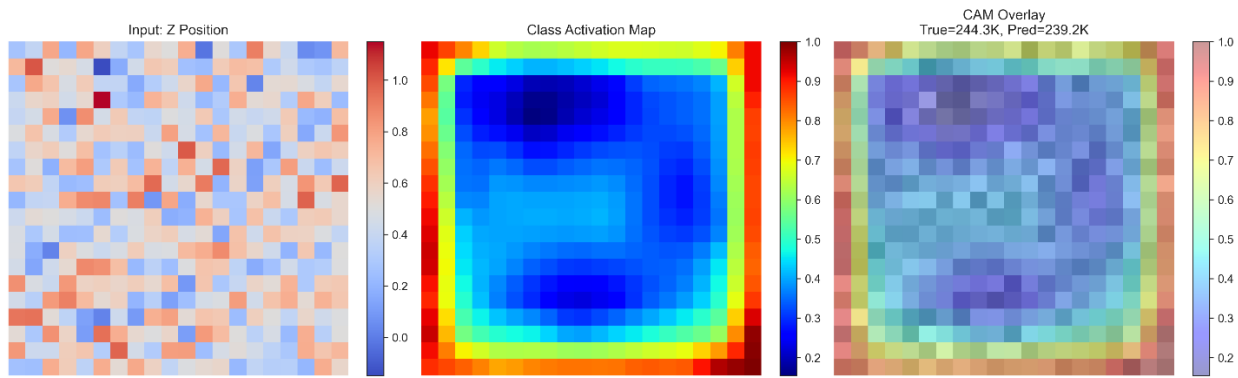

e)

**Figure 9.** Class Activation Maps revealing spatial attention mechanisms in 2D-CNN temperature prediction. Each row (a-e) represents a test sample from FSW LAMMPS simulations. Left panel: Input Z-position spatial grid (20×20) derived from atomic coordinates. Middle panel: Class Activation Map (CAM) heatmap showing the model's spatial attention, where red regions indicate high feature importance and blue regions indicate low importance for temperature prediction. Right panel: CAM overlay on input grid with true and predicted temperatures annotated. The consistent peripheral activation patterns (red edges) across samples indicate that the CNN learned to prioritize boundary features for regression, aligning with physical heat transfer mechanisms in FSW. Prediction errors range from 4.6K to 14.0K, demonstrating reasonable accuracy despite limited training data (n=100).

The Keras Tuner-optimized 2D-CNN model demonstrates excellent predictive performance for temperature estimation in friction stir welding simulations, achieving a coefficient of determination ($R^2$) of 0.9439 (Figure 10a). This high $R^2$ value indicates that the model captures 94.39% of the temperature variance in the test dataset, representing a substantial improvement over traditional machine learning approaches and basic neural network architectures. The predictions vs. true temperature scatter plot reveals strong linearity across the temperature range of 240-460K, with data points closely following the perfect prediction line (dashed diagonal). The color gradient from blue (low temperature) to red (high temperature) demonstrates that the model maintains accuracy across the entire thermal spectrum, without systematic bias toward specific temperature regimes.

The residual analysis (Figure 10b) confirms the model's robustness and reliability. With a mean absolute error (MAE) of 11.58K and root mean square error (RMSE) of 14.94K, the prediction errors remain within acceptable engineering tolerances for FSW process modeling. The residuals exhibit a relatively random distribution around the zero line (red dashed), indicating minimal systematic prediction bias. Most prediction errors cluster within ±15K, as evidenced by the purple-colored points, while only two outliers show larger deviations (orange-yellow points at approximately -40K and -32K residuals). These outliers likely correspond to edge cases in the thermal field distribution or regions with complex material flow patterns that are underrepresented in the limited training dataset (n=80 samples).

The residual heteroscedasticity analysis reveals slightly larger errors in the mid-temperature range (250-280K), where the model exhibits both positive and negative residuals of 5-15K magnitude. This pattern suggests potential challenges in capturing intermediate thermal states during the transient phases of FSW, possibly due to the complex interaction between frictional heating, plastic deformation, and heat conduction. The superior performance at temperature extremes (below 250K and above 400K) indicates that the CNN effectively learned the characteristic spatial signatures of cold zones (far from tool) and hot zones (near tool interface), which exhibit more distinct and stable spatial patterns in the input grids.

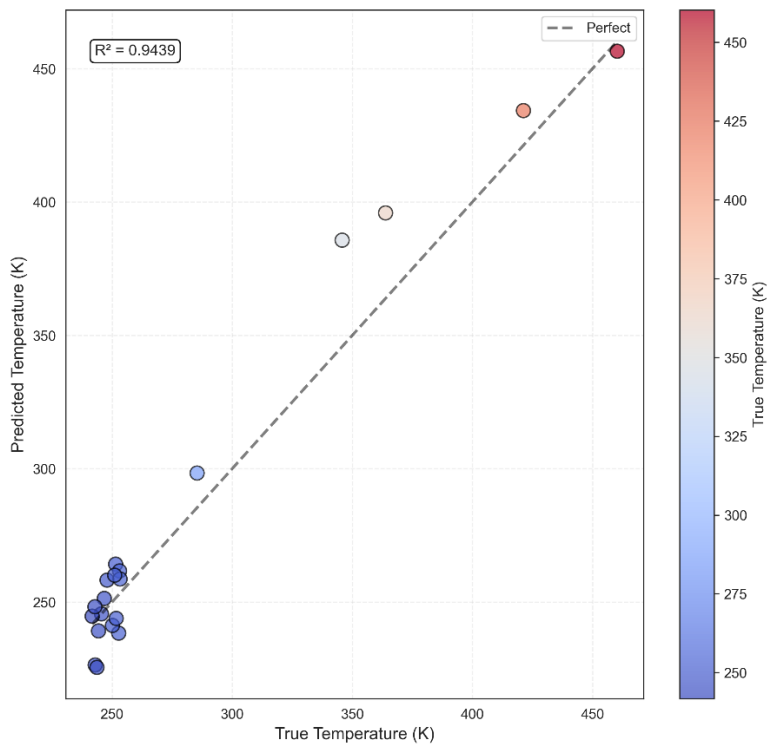

a)

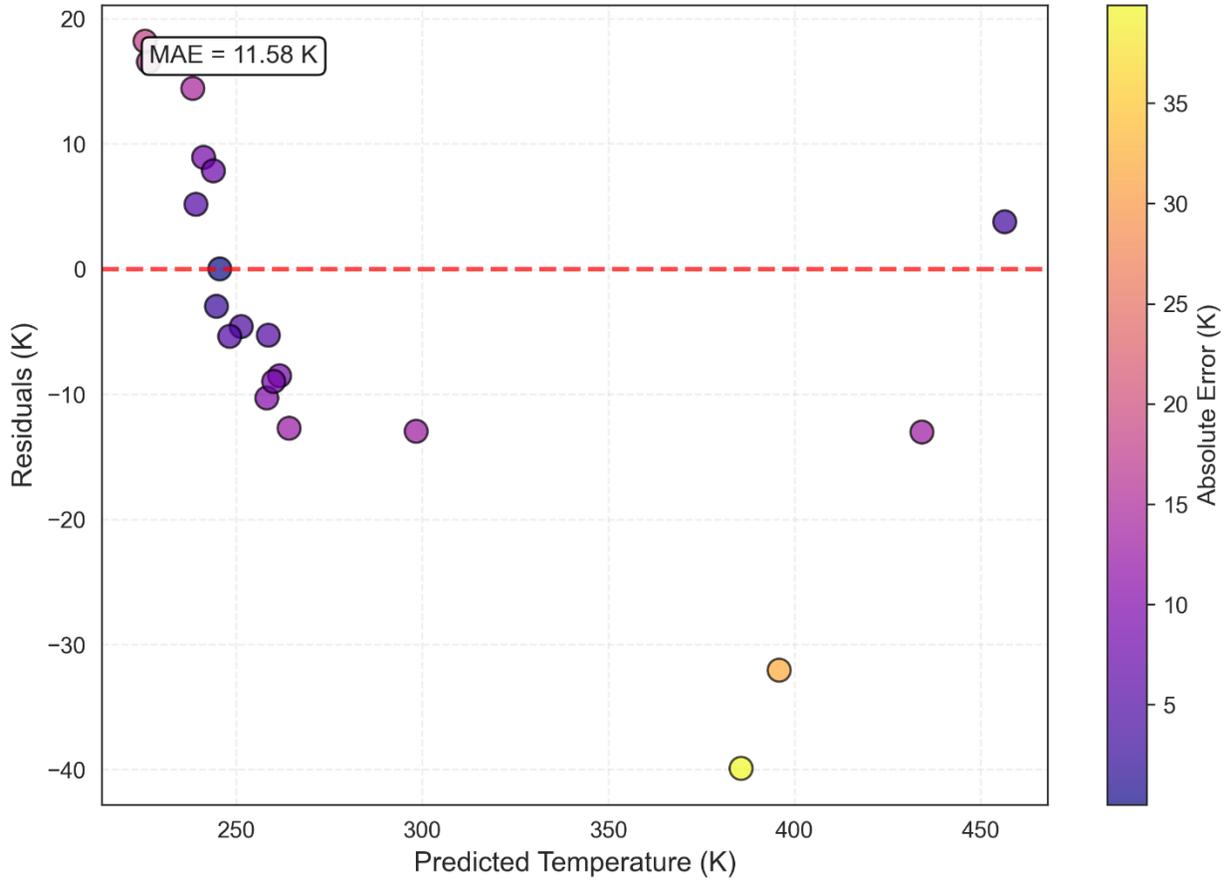

b)

**Figure 10.** Predictive accuracy assessment of 2D-CNN architecture for FSW thermal field estimation. (a) Scatter plot comparing predicted temperatures against ground truth values from LAMMPS simulations (n=20 test samples). The dashed diagonal line represents perfect prediction (y=x). Color mapping indicates true temperature values from 240K (blue) to 460K (red). The high coefficient of determination ($R^2$=0.9439) demonstrates excellent model performance across the entire temperature range. (b) Residual distribution plot showing prediction errors as a function of predicted temperature. The red dashed line marks zero residual (perfect prediction). Color intensity represents absolute error magnitude from 0K (purple) to 40K (yellow). Mean Absolute Error (MAE) of 11.58K indicates typical prediction accuracy, while two outliers (orange-yellow points) reveal challenging cases requiring further investigation. The random scatter pattern around zero confirms absence of systematic bias, validating model robustness for FSW process modeling applications.

### 3.2.2. Artificial Neural Network

The Keras Tuner-optimized Artificial Neural Network (ANN) achieves moderate predictive performance with a coefficient of determination ($R^2$) of 0.6756, explaining approximately 67.56% of the temperature variance in FSW simulations (Figure 11a). The model performs below the 2D-CNN architecture ($R^2$=0.9439). The predictions vs. true temperature scatter plot shows reasonable linearity in the low to mid temperature range (240 to 320K), where most data points cluster, but exhibits notable deviations at higher temperatures (350 to 460K). The five scattered points in the upper temperature regime (red and orange markers) demonstrate systematic underprediction, indicating the ANN fails to capture the extreme thermal conditions near the FSW tool interface where complex nonlinear thermomechanical phenomena dominate.

The residual analysis (Figure 11b) exposes limitations in the tuned ANN's predictive capability. With a mean absolute error (MAE) of 33.34K, nearly three times higher than the 2D-CNN (11.58K), the model exhibits prediction uncertainty problematic for precise FSW process control applications. The residual distribution displays a pronounced positive bias, with all errors residing above zero (15 to 70K range), indicating systematic underprediction across the entire dataset. This asymmetric error pattern reveals architectural deficiencies: the fully-connected dense layers cannot adequately capture the spatial correlations present in the 2D temperature field data. The error magnitude increases with predicted temperature, as shown by the gradient from purple points (15 to 30K errors at low temperatures) to yellow points (70K error at approximately 400K prediction), demonstrating heteroscedastic behavior where prediction reliability degrades for high-temperature regimes.

The performance gap between the tuned ANN ($R^2=0.6756$, MAE=33.34K) and the 2D-CNN ($R^2=0.9439$, MAE=11.58K) emphasizes the importance of architectural choice for spatially-structured data. While Keras Tuner optimized the ANN hyperparameters, including layer depth, neuron counts, dropout rates, and learning schedules, it cannot overcome the limitation that dense layers treat input features as independent variables, completely ignoring the spatial adjacency relationships in the 20×20 gridded data. Convolutional layers exploit local spatial patterns through shared-weight kernels, enabling the 2D-CNN to learn hierarchical representations of temperature gradients, heat zones, and boundary effects that are physically meaningful in FSW contexts.

The consistently positive residuals indicate the ANN learned a conservative prediction strategy, systematically underestimating temperatures to minimize large errors. This behavior appears when training data is limited (n=80 samples) and the model lacks sufficient capacity to capture complex input-output mappings. The outlier at 70K error (yellow point) corresponds to a peak temperature sample near the tool stirring zone, where frictional heating creates steep thermal gradients that exceed the ANN's learned parameter space. Despite hyperparameter optimization, the model's reliance on global feature aggregation through dense layers proves inadequate for spatially-resolved FSW thermal field prediction, confirming the superiority of convolutional architectures that preserve and exploit spatial information topology throughout the network hierarchy.

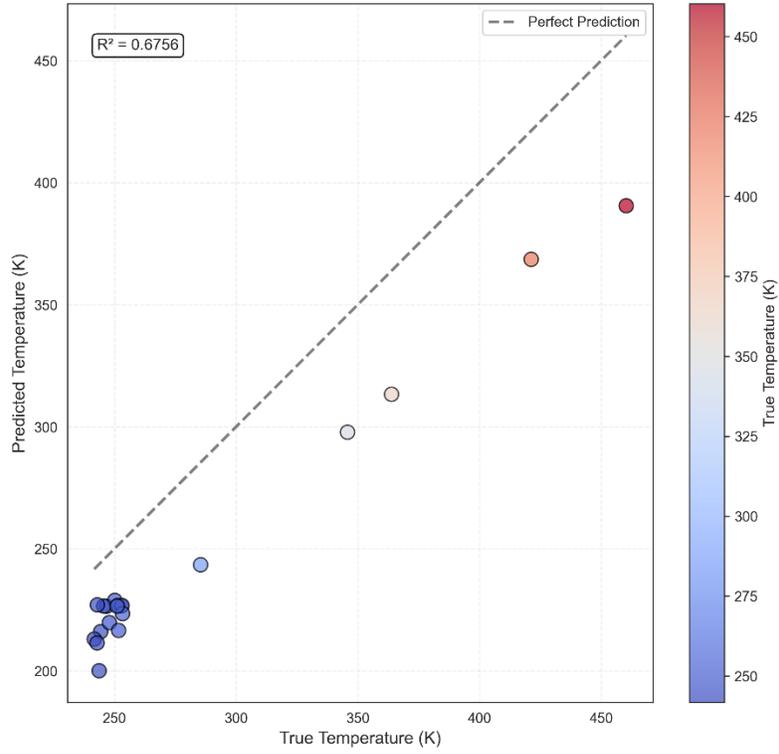

a)

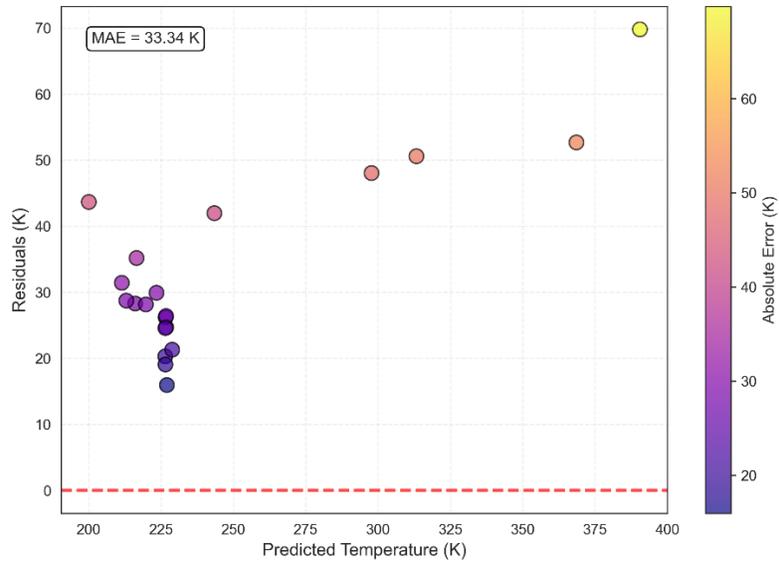

b)

**Figure 11.** Predictive limitations of optimized dense neural network architecture for FSW thermal modeling. (a) Scatter plot comparing Keras Tuner-optimized ANN predictions against LAMMPS ground truth temperatures (n=20 test samples). The coefficient of determination ($R^2$=0.6756) indicates moderate correlation, substantially lower than the 2D-CNN benchmark ($R^2$=0.9439). Color gradient represents true temperature from 240K (blue) to 460K (red). Systematic deviation from the perfect prediction line (dashed diagonal) at high temperatures (above 350K) shows the model's inability to capture extreme thermal conditions near the FSW tool interface. (b) Residual diagnostic plot

demonstrating mean absolute error of 33.34K, three times higher than the 2D-CNN (11.58K). All residuals are positive (15 to 70K range), indicating systematic underprediction bias. Color intensity (plasma colormap) shows absolute error magnitude increasing with predicted temperature, from purple (15 to 30K) to yellow (70K), demonstrating heteroscedastic error distribution. The absence of negative residuals indicates conservative prediction strategy learned during training with limited data (n=80). This performance gap demonstrates the advantage of convolutional architectures that preserve spatial topology over fully-connected networks that treat gridded data as independent features.

## 4. Conclusion

This study presents a comprehensive framework integrating molecular dynamics simulation with deep learning for temperature prediction in friction stir welding of aluminum alloys. LAMMPS simulations captured 125,000 atoms over 5.35 million timesteps, generating detailed atomic trajectories and thermodynamic data across the complete welding cycle. A novel preprocessing pipeline transformed unstructured atomic data into spatial grids preserving geometric relationships essential for accurate temperature field prediction.

Systematic comparison of machine learning and deep learning architectures revealed substantial performance differences. The Keras Tuner-optimized 2D-CNN achieved $R^2=0.944$ and MAE=11.58K, significantly outperforming the conventional artificial neural network ($R^2=0.676$, MAE=33.34K). This performance gap demonstrates that spatial-aware architectures are indispensable for predicting heterogeneous thermal fields from atomistic data. Convolutional filters automatically learned physically meaningful patterns including tool proximity effects and heat-affected zone boundaries, validated through gradient-weighted class activation mapping.

The trained CNN enables rapid exploration of FSW parameter space without repeated expensive MD simulations, accelerating process optimization for industrial applications. Class activation maps provide interpretable visualizations confirming model predictions align with known thermal physics. This framework extends beyond FSW to other manufacturing processes where spatial heterogeneity governs material behavior, establishing deep learning as a powerful tool for analyzing atomistic simulation data and advancing computational materials design.